\documentclass[fleqn, usenatbib]{mnras}

\usepackage{newtxtext, newtxmath}
\usepackage[T1]{fontenc}
\usepackage{ae, aecompl}

\usepackage{graphicx} % Figures
\usepackage[usenames]{color} % Coloured text

\graphicspath{{figs/}}

\definecolor{green}{rgb}{0.3, 0.8, 0.0}

\newcommand{\Rstar}{\ensuremath{R_\star}}

\newcommand{\Mstar}{\ensuremath{M_\star}}

\newcommand{\Rsun}{\ensuremath{R_{\sun}}}

\newcommand{\Msun}{\ensuremath{M_{\sun}}}

\newcommand{\Mjup}{\ensuremath{M_{\text{jup}}}}

\newcommand{\Rjup}{\ensuremath{R_{\text{jup}}}}

\newcommand{\Mp}{\ensuremath{M_{\text{p}}}}

\newcommand{\Rp}{\ensuremath{R_{\text{p}}}}

\newcommand{\Bp}{\ensuremath{B_{\text{p}}}}

\newcommand{\fpl}{\ensuremath{f_{\text{p}}}}

\newcommand{\fc}{\ensuremath{f_{\text{c}}}}

\newcommand{\Rm}{\ensuremath{R_{\text{m}}}}

\title[The radio environment of HD189733b]{MOVES II. Tuning in to the radio environment of HD189733b}

\author[R.~D.~Kavanagh et al.]{R.~D.~Kavanagh$^{1}$\thanks{E-mail: \texttt{kavanar5@tcd.ie}}, 
A.~A.~Vidotto$^{1}$, 
D.~\'{O}~Fionnag\'{a}in$^{1}$,
V.~Bourrier$^{2}$,
R.~Fares$^{3,4}$, \newauthor
M.~Jardine$^{5}$,
Ch.~Helling$^{6}$,
C.~Moutou$^{7}$,
J.~Llama$^{8}$,
P.~J.~Wheatley$^{9}$ \\
$^{1}$School of Physics, Trinity College Dublin, The University of Dublin, Dublin 2, Ireland \\
$^{2}$Observatoire de l'Universit\'e de Gen\'eve, Chemin des Maillettes 51, Versoix, CH-1290, Switzerland \\
$^{3}$Physics Department, United Arab Emirates University, P.O.~Box 15551, Al-Ain, United Arab Emirates \\
$^{4}$University of Southern Queensland, Centre for Astrophysics, Toowoomba, Queensland, 4350, Australia \\
$^{5}$SUPA, School of Physics and Astronomy, University of St Andrews, North Haugh, St Andrews, Fife, Scotland, KY16 9SS \\
$^{6}$Centre for Exoplanet Science, University of St Andrews, St Andrews KY16 9SS, UK \\
$^{7}$CNRS/CFHT, 65-1238 Mamalahoa Highway, Kamuela HI 96743, USA \\
$^{8}$Lowell Observatory, 1400 W.~Mars Hill Rd, Flagstaff.~AZ 86001.~USA \\
$^{9}$Department of Physics, University of Warwick, Coventry CV4 7AL, UK \\
}

\date{Accepted XXX. Received YYY; in original form ZZZ}

\pubyear{2018}

\begin{document}
\label{firstpage}
\pagerange{\pageref{firstpage}--\pageref{lastpage}}
\maketitle

% #########################################################
%  Abstract
% #########################################################

\begin{abstract}
We present stellar wind modelling of the hot Jupiter host HD189733, and predict radio emission from the stellar wind and the planet, the latter arising from the interaction of the stellar wind with the planetary magnetosphere. Our stellar wind models incorporate surface stellar magnetic field maps at the epochs Jun/Jul~2013, Sep~2014, and Jul~2015 as boundary conditions. We find that the mass-loss rate, angular momentum-loss rate, and open magnetic flux of HD189733 vary by 9\%, 40\%, and 19\% over these three epochs. Solving the equations of radiative transfer, we find that from 10~MHz--100~GHz the stellar wind emits fluxes in the range of $10^{-3}$--$5$~$\mu$Jy, and becomes optically thin above 10~GHz. Our planetary radio emission model uses the radiometric Bode's law, and neglects the presence of a planetary atmosphere. For assumed planetary magnetic fields of 1--10~G, we estimate that the planet emits at frequencies of 2--25~MHz, with peak flux densities of $\sim10^2$~mJy. We find that the planet orbits through regions of the stellar wind that are optically thick to the emitted frequency from the planet. As a result, unattenuated planetary radio emission can only propagate out of the system and reach the observer for 67\% of the orbit for a 10~G planetary field, corresponding to when the planet is approaching and leaving primary transit. We also find that the plasma frequency of the stellar wind is too high to allow propagation of the planetary radio emission below 21~MHz. This means a planetary field of at least 8~G is required to produce detectable radio emission.
\end{abstract}

% Keywords

% #########################################################
%	Introduction
% #########################################################

\section{Introduction}

Low-mass stars ($0.1-1.3$ \Msun) lose mass throughout their entire lives in the form of quiescent stellar winds. These stars also exhibit magnetic activity at their surfaces, and are observed to undergo variations in magnetic field strength \citep[e.g.][]{fares17} and polarity reversals over time \citep[e.g.][]{donati08, borosaikia16}. The Zeeman Doppler imaging (ZDI) technique, used to reconstruct large-scale surface stellar magnetic fields \citep{donati97}, has revealed that magnetically active stars exhibit complex magnetic topologies that can vary as rapidly as a few rotation periods \citep{borosaikia15}, over one magnetic cycle \citep{fares09, borosaikia16}, and over evolutionary timescales \citep{petit08, vidotto14b}. These variations in field strength, both over time and across the stellar surface, drive stellar wind outflows that vary in strength over different timescales \citep{nicholson16}. 

As the stellar wind propagates out into the interplanetary environment, the variations of its properties are most strongly felt by close-in exoplanets (orbital distances $a$ < 0.5 au). These exoplanets form a large proportion of the detected planets orbiting low-mass stars. At such close proximity, these planets are likely to be subjected to much harsher wind conditions from the host star than those experienced by the solar system planets \citep{vidotto15}. For example, close-in hot Jupiters have been observed to exhibit extended atmospheres, which is shaped by the interaction of the stellar wind with the planetary atmosphere \citep{lecavelier12, bourrier13, bourrier16}. The presence of close-in planets may also enhance the magnetic activity of the host star, both through tidal and magnetic interactions \citep{cuntz00, ip04, shkolnik08}. 

Stellar winds are believed to power exoplanetary radio emission, similarly to what occurs in the solar system for Earth, Jupiter, Saturn, Uranus, and Neptune. This radio emission occurs via the Cyclotron Maser instability \citep[see][]{treumann06}, wherein energetic electrons spiral through planetary magnetic field lines towards polar regions of strong magnetic field. This emission process has also been observed for low-mass stars \citep{bingham01, llama18}. The process for planets is powered by the dissipation of the magnetic flux of the stellar wind on the magnetosphere of the planet. As close-in exoplanets are likely to be subjected to more energetic winds, it is expected that radio emission occurs for these planets at higher powers \citep{zarka01}. If detectable, this would serve as a new direct detection method for exoplanets, show that exoplanets are magnetised, and also act as a method to probe the stellar wind of the host star. So far, there has yet to be a conclusive detection of exoplanetary radio emission, despite numerous efforts \citep{smith09, lazio10, lecavelier13, sirothia14, ogorman18}.

However, the detection of planetary radio emission could be inhibited due to dense stellar winds, which can absorb emission at radio wavelengths \citep{vidotto17a}. The winds of low-mass stars are known to be sources of radio emission, arising through thermal free-free processes \citep{panagia75, wright75, gudel02, ofionnagain19}, which depend on both the density and temperature of the wind. By the same process, the wind can self-absorb the generated free-free radio emission if the optical depth is high enough. As a result, if a planet emitting its own radio emission orbits through regions of the wind that are optically thick to the planetary frequency, the planetary emission will also be absorbed.

In this paper, we investigate the radio environment of HD189733b, which orbits the main-sequence K2V star HD189733. As part of the MOVES (Multiwavelength Observations of an eVaporating Exoplanet and its Star, PI: V.~Bourrier) collaboration, we model the wind of the star using surface magnetic field maps presented in \cite{fares17}, and calculate the radio emission from the stellar wind and planetary magnetosphere. Due to the close proximity of the planet to its host star, the system has been the subject of many star-planet interaction studies \citep{fares10, cauley18}. The host star is also very active. Its magnetic field has been observed to vary in unsigned average field strength from 18 to 42~G over a 9 year period \citep{fares17}. This short-term variation ($\sim1$ year), in combination with the small orbital distance of the planet and strong stellar magnetic field, is likely to power time-varying radio emission from the planet that could be orders of magnitudes larger than those from planets in the solar system. The variations of the stellar wind properties at the planetary orbit could also lead to variations in the UV transit lightcurve over time \citep{lecavelier12, bourrier13, llama13}.

The layout of this paper is as follows: in Section~\ref{sec:stellar wind}, we present our modelling and radio emission calculation of the wind of HD189733. Then, in Section~\ref{sec:planet radio emission} we calculate the planetary radio emission. We investigate different scenarios where the planetary radio emission may be unable to propagate out of the planetary system in Section~\ref{sec:propagation of the planetary radio emission}. We discuss our findings in Section~\ref{sec:discussion}, and then present a summary and conclusions in Section~\ref{sec:conclusions}.

% Table of stellar and planetary parameters
\begin{table}
\caption{Planetary and stellar parameters of HD189733b and its host star.}
\label{table:HD189733 parameters}
\centering
\begin{tabular}{llc}
\hline
Parameter & Value & Reference \\
\hline
\textbf{Star:} \\
Stellar mass (\Mstar) & 0.78 \Msun & 1 \\
Stellar radius (\Rstar) & 0.76 \Rsun & 1 \\
Rotational period ($P_\text{rot}$) & 11.94 days & 2 \\
Distance ($d$) & 19.8 pc & 3 \\
\textbf{Planet:} \\
Planetary mass (\Mp) & 1.13 \Mjup & 4 \\
Planetary radius (\Rp) & 1.13 \Rjup & 4 \\
Orbital distance ($a$) & 8.8 \Rstar & 1 \\
Orbital period ($P_\text{orb}$) & 2.2 days & 1 \\
\hline
\multicolumn{3}{p{0.9\columnwidth}}{1: \cite{llama13}, 2: \cite{fares10}, 3: \cite{gaiaDR2}, 4: \cite{2017AJ....153..136S}}
\end{tabular}
\end{table}

% #########################################################
%  Stellar wind of HD189733
% #########################################################

\section{Stellar wind of HD189733}
\label{sec:stellar wind}

% #########################################################
%	Wind modelling
% #########################################################

\subsection{Modelling the stellar wind}
\label{sec:wind modelling}

To model the wind of HD189733, we use the 3D magnetohydrodynamical (MHD) code BATS-R-US \citep{powell99, toth12}, modified by \cite{vidotto12}. BATS-R-US has been used to model stellar wind outflows in the context of low-mass and planet-hosting stars \citep[e.g.][]{vidotto12, llama13, cohen14, alvaradogomez16, ofionnagain19}.

BATS-R-US solves the ideal set of MHD equations, for conservation of mass, magnetic flux, momentum, and energy respectively:
%
% Mass continuity
\begin{equation}
\frac{\partial}{\partial t} \rho + \nabla \cdot (\rho\boldsymbol{u}) = 0,
\end{equation}
%
% Conservation of magnetic flux
\begin{equation}
\frac{\partial}{\partial t} \boldsymbol{B} + \nabla \cdot (\boldsymbol{uB} - \boldsymbol{Bu}) = 0,
\end{equation}
%
% Conservation of momentum
\begin{equation}
\frac{\partial}{\partial t} (\rho \boldsymbol{u}) + \nabla \cdot \bigg[ \rho\boldsymbol{uu} + \bigg( p + \frac{B^2}{8\pi} \bigg) \boldsymbol{I} - \frac{\boldsymbol{BB}}{4\pi} \bigg] = \rho \boldsymbol{g},
\end{equation}
%
% Conservation of energy
\begin{equation}
\frac{\partial}{\partial t} \varepsilon + \nabla \cdot \bigg[ \bigg( \varepsilon + p + \frac{B^2}{8\pi} \bigg) \boldsymbol{u} - \frac{(\boldsymbol{u} \cdot \boldsymbol{B}) \boldsymbol{B}}{4\pi} \bigg] = \rho \boldsymbol{g} \cdot \boldsymbol{u}.
\end{equation}
These equations are solved for a plasma with a mass density $\rho$, velocity $\boldsymbol{u} = \{ u_x, u_y, u_z \}$, magnetic field $\boldsymbol{B} = \{ B_x, B_y, B_z \}$, and thermal pressure $p$. Here, $\boldsymbol{I}$ is the identity matrix, and $\varepsilon$ is the total energy density:
%
% Total energy density
\begin{equation}
\varepsilon = \frac{\rho u^2}{2} + \frac{p}{\gamma - 1} + \frac{B^2}{8\pi},
\end{equation}
where $\gamma$ is the polytropic index. 

The inputs of BATS-R-US are the stellar mass $M_\star$ and radius \Rstar, which provide the surface gravity $\boldsymbol{g}$, coronal base temperature $T_0$ and number density $n_0$, rotational period $P_\text{rot}$, and surface magnetic field map. In our wind simulations, we make the following assumptions:
%
% List of assumptions
\begin{enumerate}
\item The wind is an ideal gas: $p = nk_\text{B}T$, where $n = \rho/(\mu m_\text{p})$ is the number density, $k_\text{B}$ is Boltzmann's constant, and $T$ is the temperature. Here, $\mu m_\text{p}$ is the mean mass per particle.
\item The wind is composed of fully ionised hydrogen, so we adopt a value of $\mu = 0.5$.
\item The wind is polytropic: $p \propto \rho^\gamma$.
\end{enumerate}
Provided that we have derived values for the stellar mass, radius, and rotational period, we are left with the following free parameters in our models: $T_0$, $n_0$, and $\gamma$. For the coronal base temperature $T_0$, we adopt a value of $2\times10^6$ K, which is typical for the coronae of K stars \citep{johnstone15}. For the base number density $n_0$, we assume $10^{10}$~cm$^{-3}$. We take $\gamma = 1.1$ for the polytropic index, which is similar to the effective adiabatic index of the solar wind \citep{vandoorsselaere11}. This set of parameters produce a mass-loss rate of $\sim3\times10^{-12}$ \Msun\ yr$^{-1}$ for HD189733, which is in the range of inferred mass-loss rates for active K stars \citep[see][]{wood04, jardine19, rodriguez19}.

\begin{figure*}
\centering
\includegraphics[width = 0.325\textwidth]{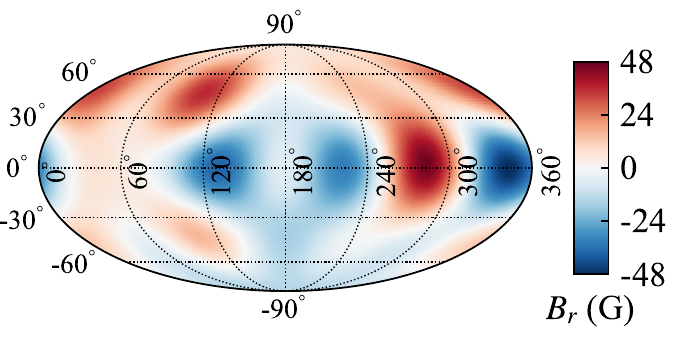}
\includegraphics[width = 0.325\textwidth]{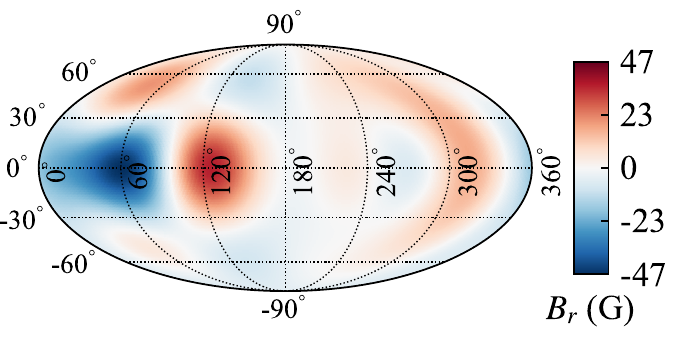}
\includegraphics[width = 0.325\textwidth]{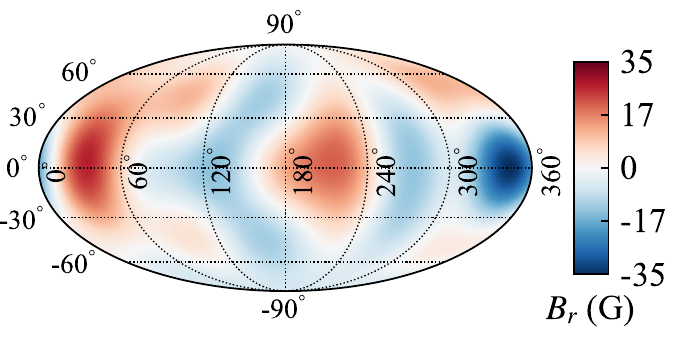}
\caption{Mollweide projections of the radial surface magnetic field of HD189733 at the three epochs Jun/Jul~2013, Sep~2014, and Jul~2015 (left to right), which were reconstructed from observations by \citet{fares17}. We implement these as boundary conditions in our stellar wind simulations.}
\label{fig:ZDI maps}
\end{figure*}

To simulate the wind variability of HD189733, we implement surface magnetic field maps obtained for the star at different epochs, which were reconstructed from observations by \citet{fares17} using the ZDI technique. The maps implemented were obtained at the epochs Jun/Jul~2013, Sep~2014, and Jul~2015, and are shown in Figure~\ref{fig:ZDI maps}. These maps allow us to estimate the yearly variability of the stellar wind properties and the planetary radio emission. The field strength at these epochs are also some of the largest observed for the star, for which we expect should drive strong radio emission from the planet. We simulate the wind with a grid that extends from $-20$~\Rstar\ to $+20$~\Rstar\ along the $x$, $y$, and $z$ axes, with the star at the centre. To solve for the magnetic field in the wind, we fix the radial component $B_r$ based on the ZDI map at each epoch, and assume an outflow or `float' condition for meridional $B_\theta$ and azimuthal $B_\phi$ components, where we take their derivative with respect to the radial coordinate to be zero. We also incorporate adaptive mesh refinement in our grid. Our grid resolution ranges from $\sim0.01$ \Rstar\ to $0.3$ \Rstar, with the finest resolution occurring in the region from 1 \Rstar\ to 2 \Rstar. This is equivalent to a refinement level of 10 quoted in \citet{nicholson16}, and has a total of 39 million cells. \citet{nicholson16} investigated the effects of increasing the grid resolution to a refinement level of 11 on the derived global wind properties, and obtained marginally more accurate results. Given that the large number of cells in a grid of this size ($\sim$ 300 million) is very computationally expensive, we adopt a refinement level of 10 in this work.

With everything in place, we set our simulation to run until a steady state is obtained. We take this to be the point where various global wind properties such as the mass-loss and angular momentum-loss rates vary by $<1\%$ between iterations. The stability of these values as a function of distance from the star also indicates that a steady state has been obtained. When these conditions are satisfied, we take the wind model to be complete.

% ZDI maps

% #########################################################
%  Wind modelling results
% #########################################################

\subsection{Stellar wind variability of HD189733}
\label{sec:wind modelling results}

From our simulations, we derive the global wind properties of HD189733 at Jun/Jul~2013, Sep~2014, and Jul~2015, as described by \citet{vidotto15}. The mass-loss rate $\dot{M}$ of the star is calculated as:
%
% Mass-loss rate calculation
\begin{equation}
\dot{M} = \oint_S \rho u_r dS,
\end{equation}
where $S$ is a spherical surface above the stellar surface at a specified distance. The angular momentum-loss rate of the star $\dot{J}$ is calculated as the flux of angular momentum through $S$:
%
% Angular momentum-loss rate calculation
\begin{equation}
\dot{J} = \oint_S \bigg[ - \frac{\varpi B_\varphi B_r}{4\pi} + \varpi u_\varphi \rho u_r \bigg] dS,
\end{equation}
where $\varpi = (x^2 + y^2)^{1/2}$ is the cylindrical radius. The unsigned open magnetic flux of the stellar wind $\Phi_\text{open}$ is:
%
% Open flux calculation
\begin{equation}
\Phi_\text{open} = \oint_S |B_r| dS.
\end{equation}

We compute $\dot{M}$, $\dot{J}$, and $\Phi_\text{open}$ at concentric spherical surfaces $S$ around the star from 10 \Rstar\ to 20 \Rstar, and take the average value in this region. Our computed values are listed in Table~\ref{table:BATSRUS results}. We find that the wind properties of HD189733 vary over time in response to its changing magnetic field topology, with the mass-loss rate, angular momentum-loss rate, and open magnetic flux varying by 9\%, 40\%, and 19\% respectively over the three modelled epochs, relative to the maximum over the three epochs. These results are comparable to those found by other stellar/solar wind models \citep{vidotto12, nicholson16, reville17}, wherein small variations in $\dot{M}$ (>20\%) and large variations in $\dot{J}$ ($\sim$50-140\%) are seen over timescales of half to multiple magnetic cycles. 

\begin{figure*}
\centering
\includegraphics[width = 0.327\textwidth]{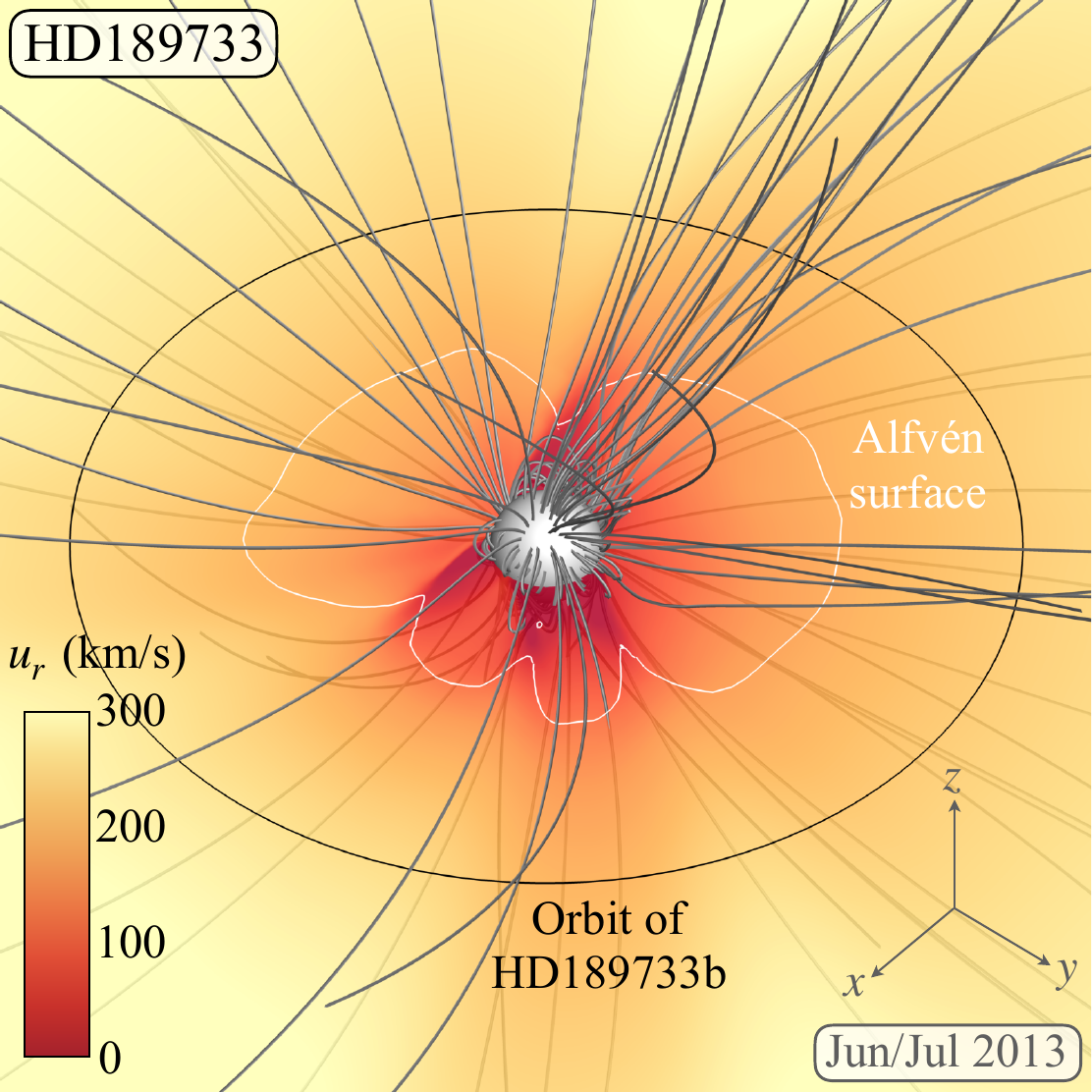}
\includegraphics[width = 0.327\textwidth]{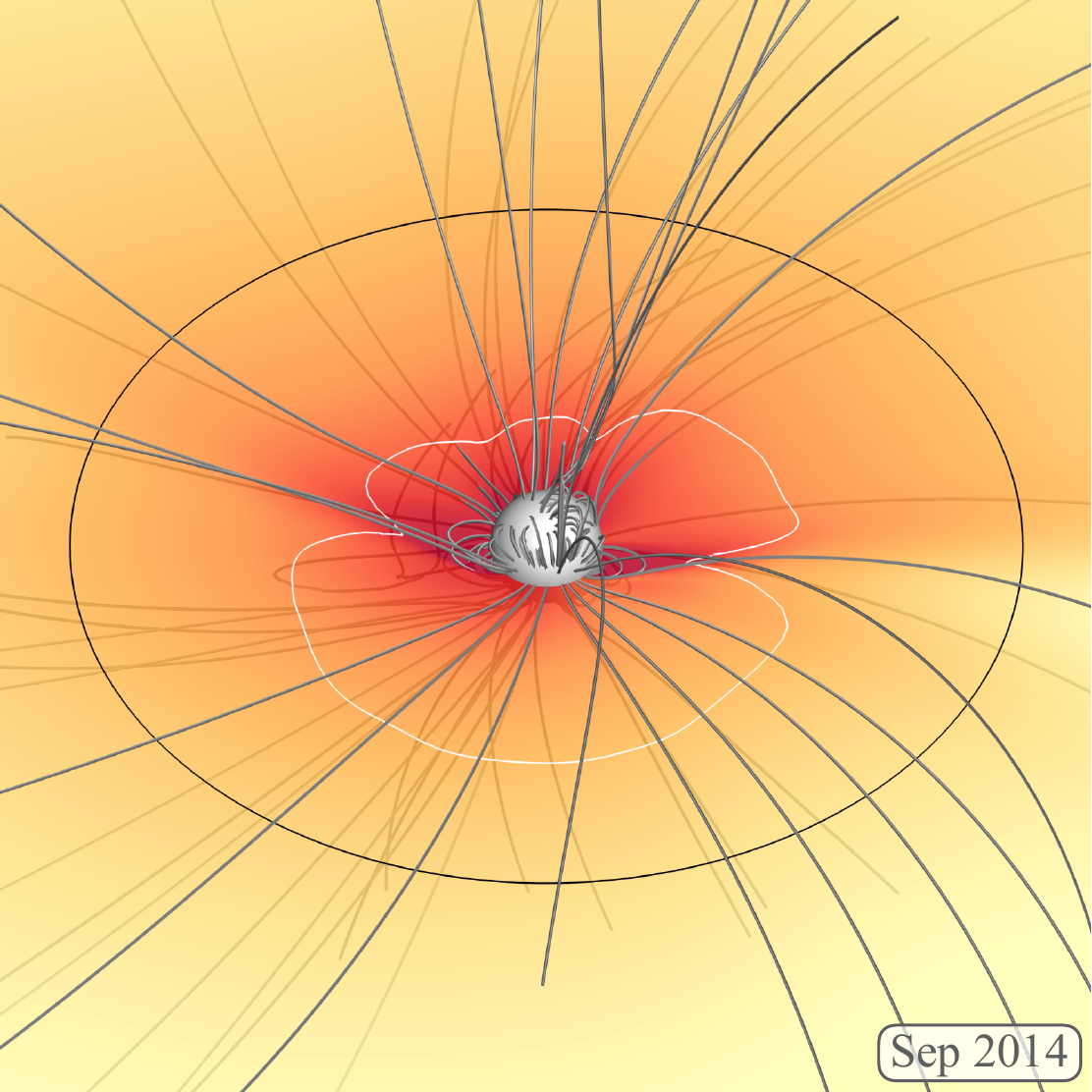}
\includegraphics[width = 0.327\textwidth]{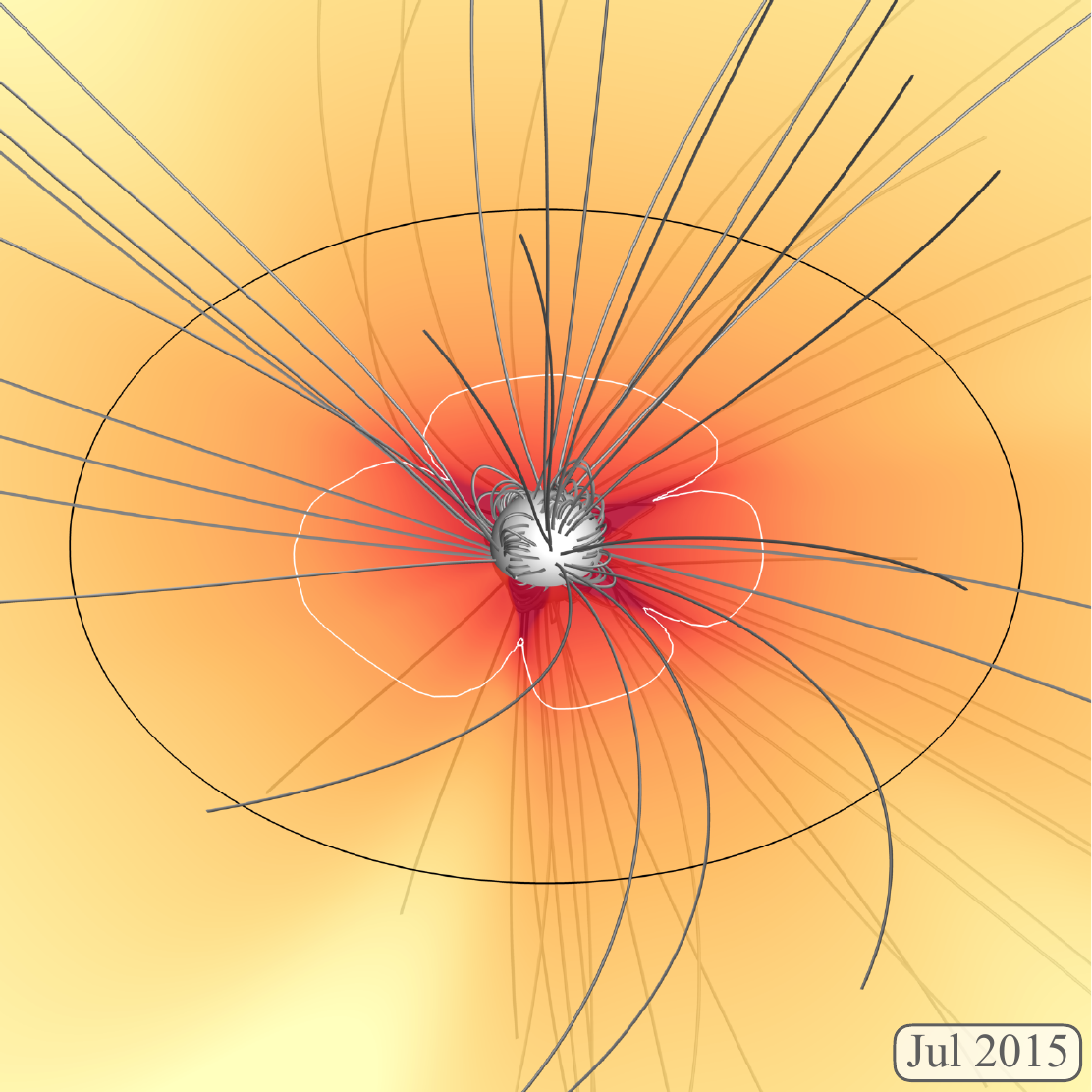}
\caption{Simulated stellar wind of HD189733, at Jun/Jul~2013, Sep~2014, and Jul~2015 (left to right). Grey lines show the large-scale structure of the magnetic field of HD189733, which is embedded in the stellar wind. Profiles of the radial velocity ($u_{r}$) of the stellar wind in the orbital plane of the planet are shown. The orbit at 8.8 \Rstar\ is shown with a black circle, and Alfv\'en surfaces are shown in white. The orientation in each panel is the same.}
\label{fig:wind contours}
\end{figure*}

As we implement magnetic field maps with complex topologies in our wind simulations, we find that the wind of HD189733 is inhomogeneous. This can be seen in the profiles of the radial wind velocity in Figure~\ref{fig:wind contours}. As a result, the planet experiences a non-uniform wind as it progresses through its orbit. Table~\ref{table:orbit wind parameters} lists the average values of various wind properties at the planetary orbit at 8.8 \Rstar, as well as their min/max values in the orbit. We find that the particle number density and velocity of the wind relative to the motion of the planet vary from 29-37\% and 25-32\% respectively throughout the orbit across the three epochs. These variations are likely to be observable in the UV transit of the planet \citep{lecavelier12, bourrier13}. From a previous transit of the planet in 2011, \citet{bourrier13} derived a stellar wind velocity of 200~km~s$^{-1}$ and density of $10^6$~cm$^{-3}$ for HD189733 at the planetary orbit, based on the Lyman-$\alpha$ absorption from the extended atmosphere of the planet in the saturation regime. These values are in good agreement with the stellar wind properties at the planetary orbit in our models. Our temperature however is a factor of $\sim30$ higher than their derived value of $3\times10^4$~K. The cause of this discrepancy could be due the fact that our models do not consider detailed heating and cooling mechanisms for the stellar wind \citep{vidotto17b}.

We also calculate the average value of the magnetic field strength at the orbit of the planet for each epoch. We find that the ambient field around the planet at the three epochs are 62.1, 55.6, and 44.6~mG respectively, which are about twice the values calculated using the potential field source surface method (PFSS) \citep{fares17}. This is due to the assumption in the PFSS method that the stellar magnetic field is in its lowest energy state, which neglects that the stellar wind stresses on the stellar magnetic field lines. From the total pressure values (ram, thermal, and magnetic pressures) listed in Table~\ref{table:orbit wind parameters}, we also see that HD189733b is subjected to a wind pressure that varies by over 2 orders of magnitude throughout its orbit and over time. 

In Figure~\ref{fig:wind contours} we also show that at each epoch the planet orbits outside of the Alfv\'en surface, where the wind poloidal velocity equals the Alfv\'en velocity: $v_\text{A} = B_\text{pol}/\sqrt{4\pi\rho}$ ($B_\text{pol}$ is the poloidal magnetic field strenght of the stellar wind). This means the planet is orbiting in the ram pressure-dominated region of the wind. As a result, we do not expect any magnetic star-planet interactions (SPI) at these epochs, as information cannot propagate back to the star \citep{shkolnik08, lanza12}. However, due to the variability of the stellar magnetic field, there may be epochs in which the planet orbits inside the Alfv\'en surface, and is in direct connection to the star. During these epochs, the planet may induce hot chromospheric spots in the star. For instance, \citet{cauley18} recently reported Ca II K modulations with a period coinciding with the planetary orbital period of 2.2 days in Aug~2013, implying that magnetic SPI may have occurred at this epoch.

% Table of wind model results
\begin{table*}
\caption{Global wind properties derived for HD189733 at different epochs. The values listed are the mass-loss rate $\dot{M}$, angular momentum-loss rate $\dot{J}$, unsigned open magnetic flux $\Phi_\text{open}$, and unsigned magnetic flux at the surface $\Phi_0$.}
\label{table:BATSRUS results}
\centering
\begin{tabular}{lcccc}
\hline
Epoch & $\dot{M}$ & $\dot{J}$ & $\Phi_\text{open}$ & $\Phi_0$ \\
& ($10^{-12} \ M_{\sun}$ yr$^{-1}$) & ($10^{31}$ erg) & ($\Phi_0$) & ($10^{23}$ Mx) \\
\hline
Jun/Jul~2013 & 3.2 & 5.5 & 0.39 & 4.5 \\
Sep~2014 & 3.0 & 5.2 & 0.48 & 3.0 \\
Jul~2015 & 2.9 & 3.3 & 0.40 & 2.7 \\
\hline
\end{tabular}
\end{table*}

% Table of average wind values at the planetary orbit
\begin{table*}
\caption{Average stellar wind parameters of HD189733 at the planetary orbit. The values listed are the particle number density $n$, velocity in the reference frame of the planet $\Delta u$, magnetic field strength $B$, temperature $T$, and total pressure $p_\text{tot}$ of the stellar wind. The values quoted in square brackets are the min/max value of each parameter throughout the orbit, respectively.}
\label{table:orbit wind parameters}
\centering
\begin{tabular}{lccccc}
\hline
Epoch & $\langle n \rangle$ & $\langle \Delta u \rangle$ & $\langle B \rangle$ & $\langle T \rangle$ & $\langle p_\text{tot} \rangle$ \\
& ($10^6$ cm$^{-3}$) & (km s$^{-1}$) & (mG) & ($10^6$ K) & ($10^{-4}$ dyn cm$^{-2}$) \\
\hline
Jun/Jul~2013 & 4.0 & 235 & 62.1 & 1.03 & 1.71 \\
& [3.3, 5.2] & [188, 270] & [2.2, 84.9] & [0.94, 1.22] & [<0.01, 2.87] \\
Sep~2014 & 3.8 & 220 & 55.6 & 0.97 & 1.37 \\
& [3.0, 4.7] & [183, 271] & [1.9, 79.1] & [0.93, 1.17] & [<0.01, 2.49] \\
Jul~2015 & 3.7 & 212 & 44.6 & 0.95 & 0.86 \\
& [3.2, 4.5] & [196, 260] & [3.9, 57.8] & [0.92, 1.05] & [0.01, 1.33] \\
\hline
\end{tabular}
\end{table*}

% Wind radial velocity contours with Alfven surfaces and orbit

% #########################################################
%   Wind radio emission
% #########################################################

\subsection{Radio emission from the stellar wind}
\label{sec:wind radio emission}

\begin{figure}
\centering
\includegraphics[width = \columnwidth]{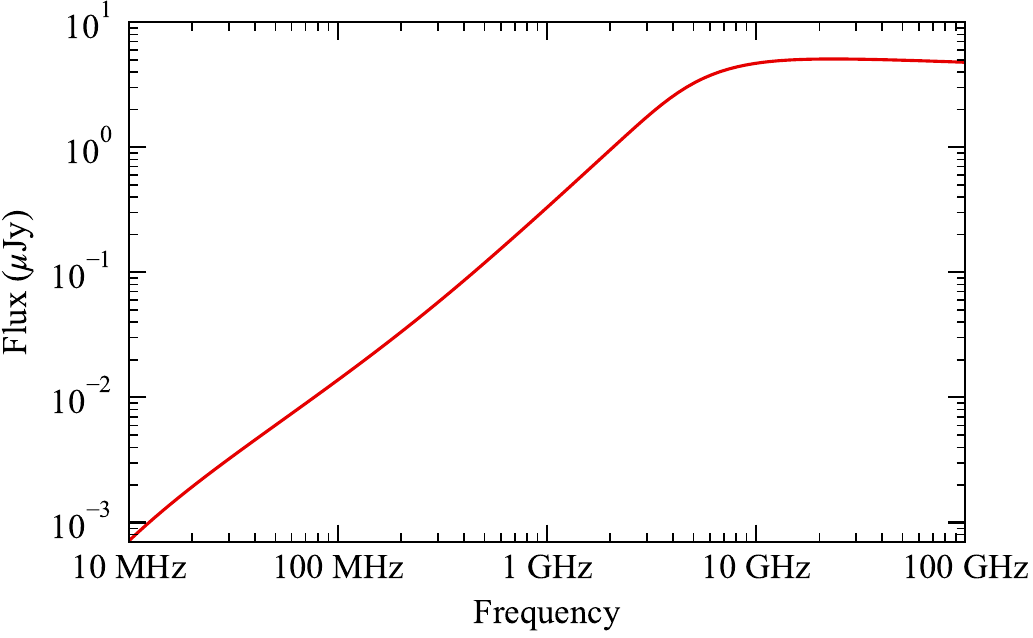}
\caption{Free-free radio emission spectra of the stellar wind of HD189733 at Jun/Jul~2013. There are negligible differences between the wind spectra for the three modelled epochs.}
\label{fig:wind radio spectrum}
\end{figure}

We calculate the free-free radio emission of the stellar wind of HD189733, implementing the numerical code developed by \citet{ofionnagain19}. In this model, we solve the equations of radiative transfer for our wind models, assuming that it emits as a blackbody. The equations are solved for the line of sight of an observer placed at $x=-\infty$, looking towards the star. The choice of the direction of the line of sight has negligible effects on the results of our calculations. We compute the free-free radio spectrum for the wind of HD189733 in the frequency range of 10 MHz to 100 GHz. In this region, we find that there are negligible differences between the fluxes for each epoch. Figure~\ref{fig:wind radio spectrum} shows the spectra calculated for Jun/Jul~2013. The flux densities calculated range from $\sim10^{-3}$ to $5$~$\mu$Jy. At such low flux densities, this emission is unlikely to be detected with current radio telescopes. However, future developments, such as the Square Kilometre Array (SKA), are likely to allow for the detection of this emission from low mass stars, such as HD189733 (see Section~\ref{sec:discussion}).

We also find that the wind of HD189733 becomes optically thin at 10~GHz. For frequencies less than this, regions of the wind are optically thick, which become larger towards lower frequencies. Inside these regions or `radio photospheres', the wind will self-absorb its own emission. In addition to this, if the planet emits cyclotron emission and orbits within the radio photosphere for the emitted frequency, this emission would also be absorbed by the stellar wind. We investigate this scenario in Section~\ref{sec:propagation - optically thick regions}. 

We note that the free-free radio emission from the stellar wind depends on the density profile of the wind, which is determined by coronal base density $n_0$, a free variable in our simulations. We discuss the effects of a lower density wind on the radio emission in Appendix \ref{sec:appendix - wind radio emission lower density}. 

% Free-free radio emission spectrum of the stellar wind

% #########################################################
%	Planetary radio emission
% #########################################################

\section{Predicting planetary radio emission from HD189733b}

% #########################################################
%   Theory
% #########################################################

\label{sec:planet radio emission}

To predict the planetary radio emission from HD189733b, we follow the description by \citet{vidotto17a} \citep[see also works by][]{zarka07}. Pressure balance between the stellar wind (ram, thermal, and magnetic) and the planet (magnetic) pressures gives us the size of the planetary magnetopause \Rm:
%
% Magnetopause size
\begin{equation}
\Rm = 2^{1/3} \Bigg[\frac{(\Bp/2)^2/8\pi}{\rho {\Delta u}^2 + p + B^2/8\pi} \Bigg]^{1/6} \Rp,
\label{eq:magnetopause size}
\end{equation}
where \Bp\ is the strength of the dipolar planetary magnetic field at the poles, \Rp\ is the planetary radius, and $\Delta u$ is the magnitude of the relative velocity between the wind $u$ and Keplerian velocity of the planet $u_\text{K}$: $\boldsymbol{\Delta u} = \boldsymbol{u} - \boldsymbol{u_\text{K}}$, $u_\text{K} = \sqrt{G\Mstar/a}$. Note that we neglect the ram and thermal components of the planetary pressure. In our model, the radio emission arises from polar cap regions at co-latitudes $\alpha_0$ from the poles:
%
% Latitude of emission region
\begin{equation}
\alpha_0 = \sin^{-1} \bigg[ {\Big(\frac{\Rp}{\Rm}\Big)}^{1/2} \bigg].
\label{eqn:alpha planet}
\end{equation}
Assuming an aligned dipolar planetary magnetic field, the field strength at this co-latitude is:
%
% Planetary magnetic field at the polar cap boundary
\begin{equation}
B(\alpha_0) = \frac{\Bp}{2} \big(1 + 3\cos^2\alpha_0 \big)^{1/2}.
\end{equation}
Radio emission then occurs via the Cyclotron Maser instability at a maximum frequency of:
%
% Cyclotron emission frequency
\begin{equation}
\fc = 2.8 \bigg( \frac{B(\alpha_0)}{1\ \text{G}} \bigg)\ \text{MHz}. 
\label{eqn:cyclotron frequency planet}
\end{equation}
The polar cap region where this emission originates forms a hollow cone with solid angle $\omega$:
%
% Solid angle of emission cone
\begin{equation}
\omega = 8\pi \sin \alpha_0 \sin \frac{\delta \alpha}{2},
\end{equation}
where $\delta \alpha$ is the thickness of the cone. We adopt a value of $\delta \alpha = 17.5^\circ$, which is the estimated thickness for Jupiter's emission cone \citep{zarka04}.

Combining all this gives us an expression for the radio flux received from the system. The magnetic power dissipated by the wind onto the magnetosphere of the planet is given by:
%
% Magnetic power dissipated by the wind
\begin{equation}
P_B = \frac{B_\perp^2}{4\pi} \Delta u \pi \Rm^2,
\label{eq:magnetic wind power}
\end{equation}
where $B_\perp$ is the magnitude of the magnetic field of the stellar wind perpendicular to the vector $\boldsymbol{\Delta u}$. In our model, we compute this as:
%
% Perpendicular B field
\begin{equation}
B_\perp = \sqrt{B^2 - \bigg( \frac{\boldsymbol{B} \cdot \boldsymbol{\Delta u} }{\Delta u} \bigg)^2 }.
\end{equation}
The flux density received by an observer at a distance $d = 19.8$~pc is then given by:
%
% Magnetic radio flux
\begin{equation}
\Phi_\text{radio} = \frac{\eta_B P_B}{d^2 \omega \fc}.
\end{equation}
Here, $\eta_B$ is the efficiency ratio for the magnetic power dissipated onto the planetary magnetosphere, which we take as the solar system value of $2\times10^{-3}$ \citep{zarka07}. This constant arises from the radiometric Bode's law. Numerical studies have found that the
radiometric Bode's law is appropriate for estimating radio emission from hot Jupiters \citep{varela18}, while some have suggested that it may overestimate the flux densities for close-in planets \citep{nichols16}. If this is indeed the case, the values we obtain here should be treated as upper limits. As we demonstrate in Section \ref{sec:propagation of the planetary radio emission}, there are several scenarios where in fact propagation of planetary radio emission may not be possible.

\begin{figure}
\centering
\includegraphics[width = \columnwidth]{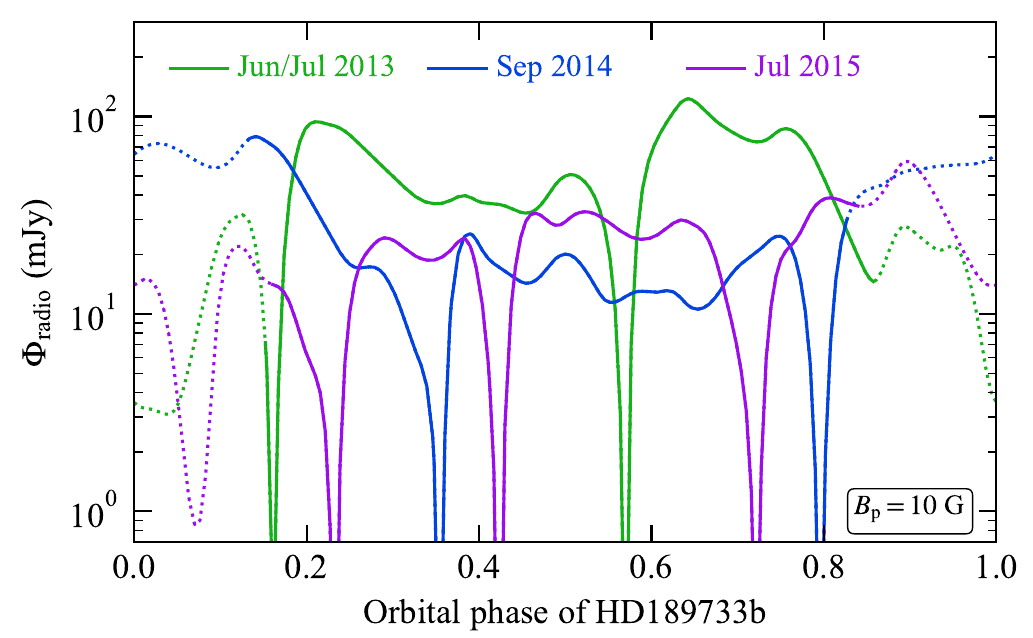}\\
\includegraphics[width = \columnwidth]{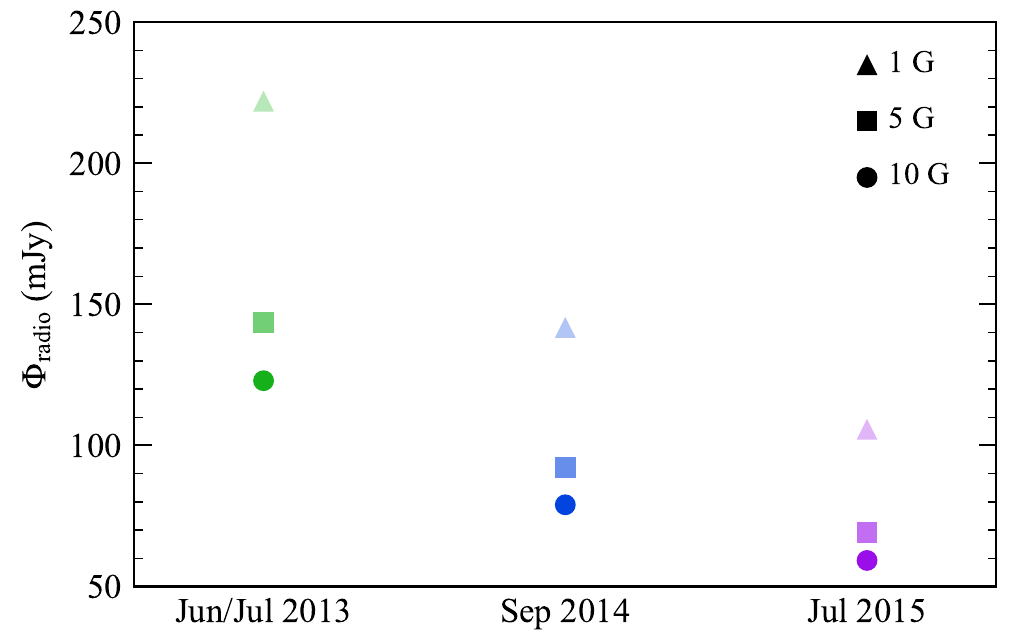}
\caption{\textit{Top}: Variations of the planetary radio flux throughout the orbit of HD189733b for the three epochs, for a field strength of \Bp\ = 10~G. The dotted segment of each line represents the region of the orbit where the planetary radio emission at 25~MHz, corresponding to a planetary field strength of 10~G, cannot propagate through the stellar wind (see Section~\ref{sec:propagation - optically thick regions}). \textit{Bottom}: Peak radio flux densities calculated for HD189733b, for planetary field strengths of 1, 5, and 10~G at each epoch.}
\label{fig:planet flux}
\end{figure}

In the above equations we use the velocity, density, pressure, and magnetic field of the stellar wind at the planetary orbit, obtained from our wind models presented in Section~\ref{sec:wind modelling results}. In our calculations, we assume planetary magnetic field strengths of \Bp\ = 1, 5, and 10~G, which covers the range of values predicted for extrasolar gas giants \citep{zaghoo18}. Table~\ref{table:planetary radio emission values} shows the calculated cyclotron frequency, magnetopause size, and co-latitude of the polar cap of HD189733b, as a function of planetary magnetic field strength. We find that these three quantities vary by small amounts throughout the orbit and between the three epochs for a given field strength. The flux densities of the radio emission however varies by around two orders of magnitude throughout the orbit. The variations of the fluxes calculated for \Bp\ = 10~G throughout the orbit of the planet, and the peak fluxes at each epoch for different planetary field strengths, are shown in Figure~\ref{fig:planet flux}. In the top panel of Figure~\ref{fig:planet flux}, we see that the flux density goes to zero at points in the orbit where $B_\perp = 0$ in the stellar wind (see Equation~\ref{eq:magnetic wind power}). We also see a general trend towards lower peak fluxes from Jun/Jul~2013 to Jul~2015 in the bottom panel of Figure~\ref{fig:planet flux}. This correlates to the trend in the average magnetic field strength of the wind at the planetary orbit shown in Table~\ref{table:orbit wind parameters}.

Comparing to recent work, \citet{zaghoo18} estimated a flux of $\sim 20$~mJy at a peak cyclotron frequency of 20~MHz for HD189733b. Our peak flux densities are $\sim5$ times larger than theirs, which is likely due to their assumption for the stellar wind power, which they extrapolated from the solar wind. As we have shown in Section~\ref{sec:wind modelling results}, the wind strength of HD189733 is much stronger than the solar wind. As a result, we expect that HD189733b receives a higher magnetic energy from the wind of its host star, and thus higher fluxes will be emitted.

Our findings tell us that while the flux density of the planetary emission is sensitive to inhomogeneities in the stellar wind, the frequency of the emission is not. The combination of these two results could allow for planetary radio emission to be easily distinguished from other sources of radio emission in the system, such as the stellar wind. The fluxes and frequencies calculated for HD189733b also place it within the detection limit of LOFAR. We discuss this further in Section~\ref{sec:discussion}.

% Table of planetary radio emission values
\begin{table}
\caption{Emitted cyclotron frequency, magnetopause size, and co-latitude of the polar cap of HD189733b, calculated for different planetary magnetic field strengths. These values only vary by small amounts through the orbit and between each epoch.}
\label{table:planetary radio emission values}
\centering
\begin{tabular}{cccc}
\hline
\Bp & \fc & \Rm & $\alpha_0$ \\
(G) & (MHz) & (\Rp) & ($\degr$) \\
\hline
1 & 2 & 1.6 & 53 \\
5 & 12 & 2.7 & 38 \\
10 & 25 & 3.3 & 33 \\
\hline
\end{tabular}
\end{table}

% Planetary radio fluxes

% #########################################################
%   Propagation of the planetary radio emission
% #########################################################

\section{Propagation of the planetary radio emission}
\label{sec:propagation of the planetary radio emission}

While we have made predictions about the planetary radio emission for different assumed magnetic field strengths, the question remains: can the emission propagate out of the planetary system? We discuss different scenarios where propagation may not be possible below.

% #########################################################
%   Propagation through the stellar wind
% #########################################################

\subsection{Free-free absorption in the stellar wind}
\label{sec:propagation - optically thick regions}

One scenario where the planetary radio emission may not propagate from the system is if the planet orbits inside the radio photosphere of the star. From the model utilised in Section~\ref{sec:wind radio emission}, we calculate the boundary of the photospheres of the stellar wind for the calculated cyclotron frequencies of 2, 12, and 25~MHz, which we chose as they correspond to planetary field strengths of 1, 5, and 10~G (see Table~\ref{table:planetary radio emission values}). The radio photosphere is defined as the point where the optical depth $\tau = 0.399$ \citep{panagia75}.

\begin{figure}
\centering
\includegraphics[width = \columnwidth]{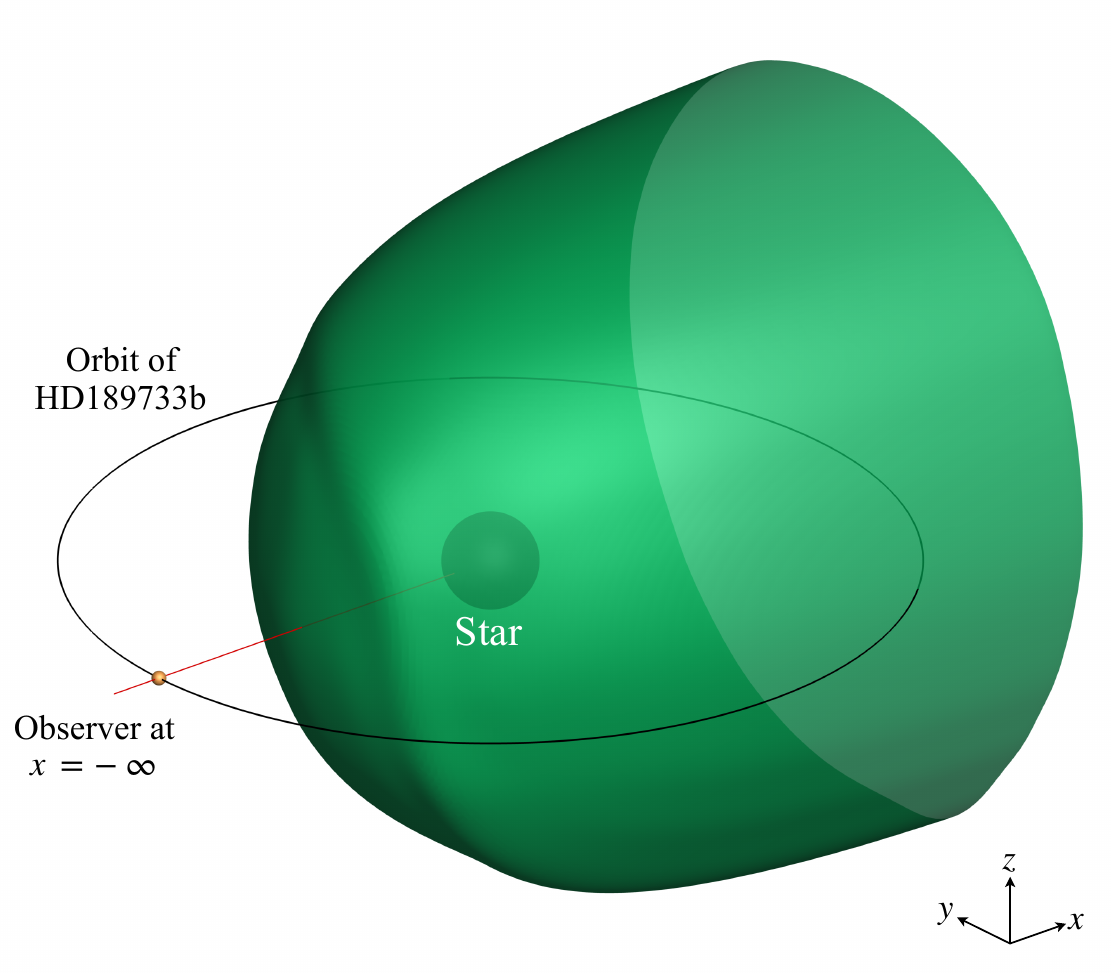}
\caption{Radio photosphere of the stellar wind of HD189733 at 25~MHz for Jun/Jul~2013, as viewed by an observer at $x=-\infty$. The region enclosed by the surface is where planetary radio emission at 25~MHz cannot propagate without being absorbed. The red line illustrates the line connecting the planet to the star seen by the observer. Note that the size of the photosphere, star, planet and its orbit are to scale here. The radio photosphere extends back to $x=10$ \Rstar.}
\label{fig:3d radio photosphere}
\end{figure}

\begin{figure}
\centering
\includegraphics[width = \columnwidth]{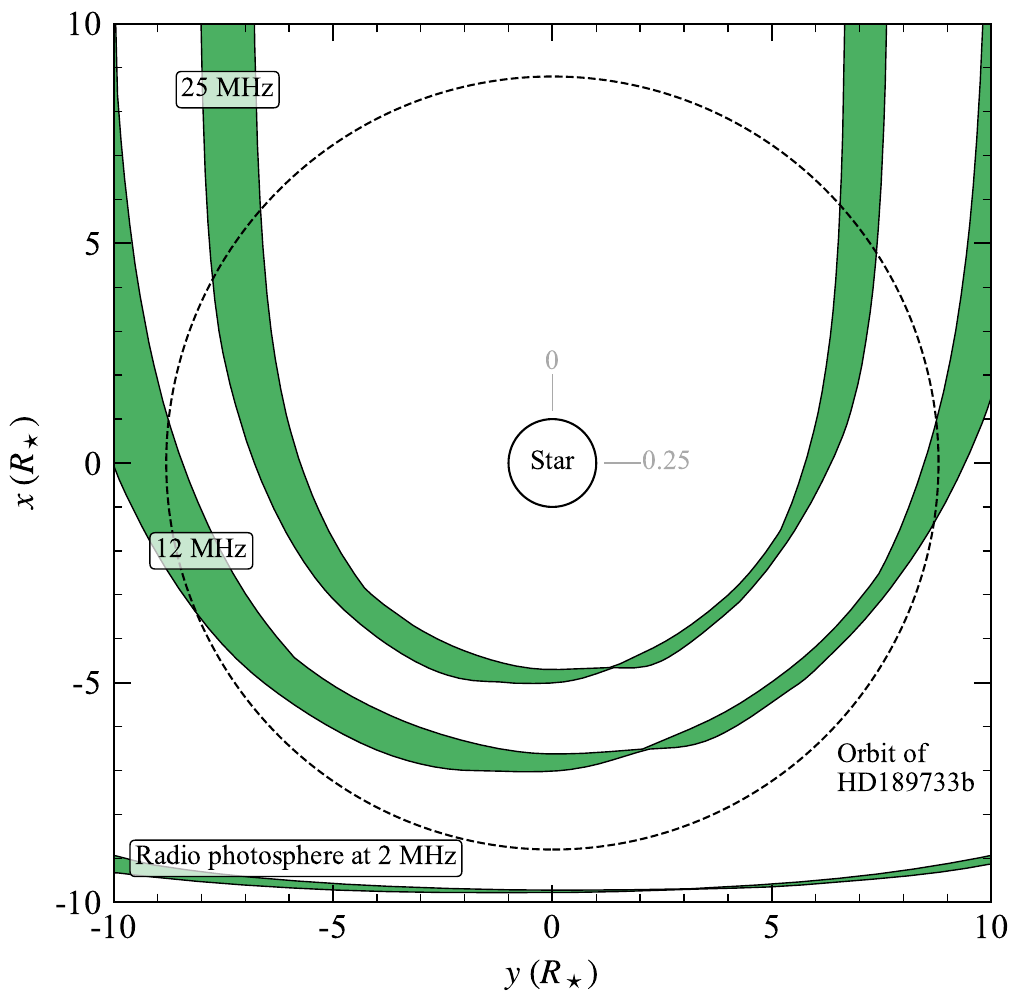}
\caption{Radio photospheres of the stellar wind at the calculated frequencies of 2, 12, and 25 MHz, in the orbital plane of the planet. The planetary orbit is shown as a black dashed circle, and the star is shown in the centre. Orbital phases of 0 and 0.25 are marked near the star. In our calculation, the observer is looking along the $x$-axis in the positive direction.}
\label{fig:radio photospheres}
\end{figure}

Figure~\ref{fig:3d radio photosphere} shows a 3D view of the radio photosphere of the wind of HD189733 at 25~MHz for Jun/Jul~2013, as viewed by an observer at $x = -\infty$. The radio photosphere is paraboloidal in shape. Inside the region enclosed by the surface shown, emission at 25~MHz will be absorbed by the stellar wind. We see that part of the planetary orbit is embedded in this region. In Figure~\ref{fig:radio photospheres}, we show the boundaries of these regions for the three epochs at 2, 12, and 25~MHz in the orbital plane of the planet. The thickness of these boundaries at each frequency arises from the different positions of the radio photosphere for the three epochs. From the boundaries, we find that the planetary radio emission can propagate for 41\% of the orbit at 12 MHz, and 67\% at 25 MHz. The planet's orbit is fully embedded in the region of the wind that is optically thick at 2 MHz. This frequency is below the ionospheric cutoff of the Earth's atmosphere. As the radio flux densities of the planet are several orders of magnitude higher than those of the stellar wind (see Figure \ref{fig:radio sensitivites}), it could be that even after the planetary emission is attenuated inside the radio photosphere of the stellar wind, a fraction of the planetary emission may still escape.

The fraction of the orbit where the emission can propagate unattenuated to an observer at Earth corresponds to when the planet is approaching and leaving primary transit. This may be a useful signature to search for in other exoplanetary radio surveys. Indeed, \citet{smith09} reported a non-detection of HD189733b during secondary transit, the region where we expect the wind to be optically thick for frequencies below 10~GHz. We note however that the frequency range of their observations correspond to planetary field strengths of $\sim100$~G, which is likely to be too strong for a hot Jupiter planet such as HD189733b \citep[see][]{zaghoo18}. It is clear however that detection of planetary radio emission is more favourable for higher frequencies, which correspond to strong planetary magnetic field strengths. We caution that the size of the radio photosphere at a specific frequency scales with the density of the stellar wind, which is a free variable in our model. The effects of varying the density on the size of these regions are discussed in Appendix \ref{sec:appendix - wind radio emission lower density}.

% 3D view of the radio photosphere

% Radio photospheres of the stellar wind

% #########################################################
%   Plasma frequency of the wind
% #########################################################

\subsection{Plasma frequency of the stellar wind}

Another process by which the planetary emission may not be able to escape the system is if the cyclotron frequency is less than the plasma frequency of the stellar wind at the planetary orbit. The plasma frequency is defined as:
%
% Plasma frequency
\begin{equation}
\fpl = 9\times10^{-3} \sqrt{\frac{n_e}{1\ \text{cm}^{-3}}}\ \text{MHz}.
\label{eqn:plasma frequency}
\end{equation}
where $n_e$ is the number density of free electrons. We take this to be equal to the number density of protons in our wind, as it is fully ionised. 

\begin{figure}
\centering
\includegraphics[width = \columnwidth]{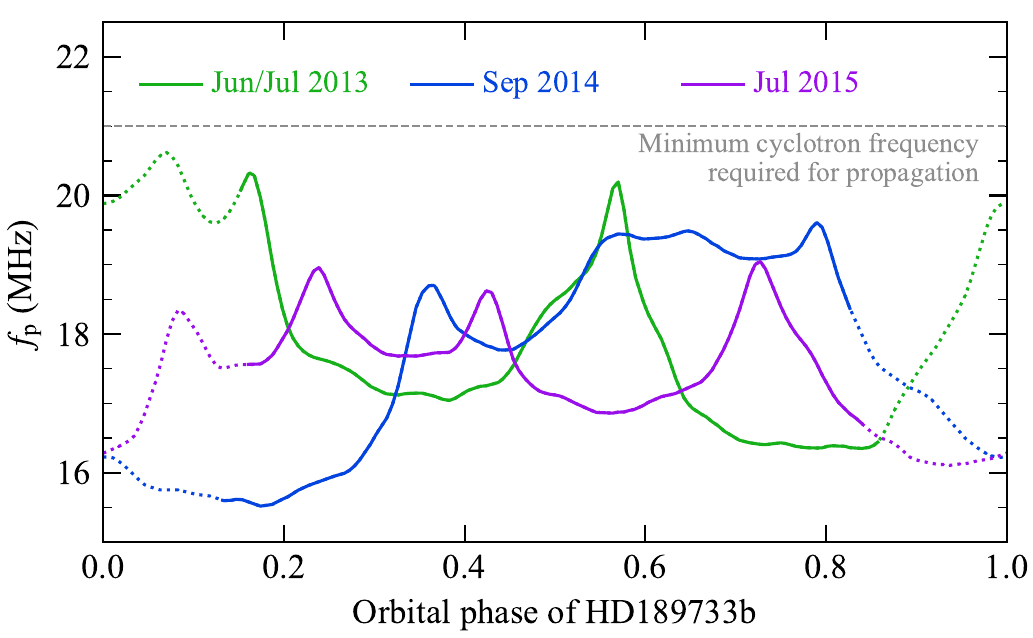}
\caption{Variations of the plasma frequency of the stellar wind at the planetary orbit for each modelled epoch. The minimum cyclotron frequency of $\sim21$~MHz required for the planetary emission to propagate through the stellar wind is marked with a grey dashed line. The dotted line segments illustrate where the planetary radio emission is absorbed by the stellar wind at 25~MHz (see Section~\ref{sec:propagation - optically thick regions}).}
\label{fig:wind plasma frequency}
\end{figure}

Neglecting the presence of the planetary atmosphere, the radio emission can propagate out from the polar emission region. However, if the emitted frequency is below the plasma frequency of the stellar wind it meets, the planetary radio emission will be reflected back and hence will not propagate out of the planetary system. This is the same process which prevents sources of radio emission below 10 MHz from penetrating the Earth's ionosphere. Figure~\ref{fig:wind plasma frequency} shows the plasma frequency of the stellar wind at the planetary orbit for the three modelled epochs. We find that only cyclotron frequencies above $\sim21$~MHz will be able to propagate through the stellar wind for the entirety of the orbit at the three modelled epochs. This requires a minimum planetary field strength of $\sim8$~G, based on the values listed in Table~\ref{table:planetary radio emission values}.

We note that the plasma frequency of the stellar wind will also scale with the wind density. For a wind that is 10 times less dense, the plasma frequency of the wind is reduced by a factor of $\sqrt{10}$, thus lowering the minimum planetary field strength required for the planetary emission to propagate to $\sim3$~G.

% Wind plasma frequency at the planetary

% #########################################################
%   Generation of the radio emission
% #########################################################

\subsection{Generation of cyclotron emission in an extended atmosphere}

Another potential issue for the detection of planetary radio emission is the planetary atmosphere itself. Recently, \citet{weber18} suggested that while close-in hot Jupiters are exposed to more energetic winds that could boost their radio emission, they also receive much higher XUV fluxes from their host stars. As a result, their atmospheres extend far out from the planet and have a higher free electron density. From Equation~\ref{eqn:plasma frequency}, this means that the plasma frequency of the atmosphere becomes very large. Therefore, a planetary field strength of hundreds of G would be required for the generated cyclotron frequency to be above the plasma frequency of the atmosphere. They suggest however that supermassive hot Jupiters may be more suitable targets for detection of planetary radio emission. As their atmospheres would be more tightly bound, the free electron density drops off more sharply with distance from the planet, and thus the radio emission can be generated above the plasma frequency of the atmosphere.

In our planetary radio emission model, we do not account for the presence of a planetary atmosphere, and so we were unable to test the predictions of \citet{weber18}. Recently however, \citet{daleyyates18} modelled the MHD equations of both the stellar and planetary winds of a solar-type star host to a hot Jupiter-type planet. They found that the generated cyclotron frequency in the planetary magnetosphere was at least a factor of 10 lower than the plasma frequency throughout the planetary atmosphere. As it is expected that the atmosphere of HD189733b is extended \citep{lecavelier12, bourrier13}, it could prevent the generation of cyclotron emission in the case of HD189733b.

% #########################################################
%  Detection potential
% #########################################################

\section{Detection potential with current and future radio telescopes}
\label{sec:discussion}

From our calculations of the radio emission from the planetary magnetosphere and the wind of the host star, it is clear that their radio signatures are very different. For the planet, it emits at a constant single frequency for a given planetary field strength, with the flux of its emission varying from $\sim1$ to $10^2$~mJy as it progresses through its orbit. The stellar wind on the other hand emits at much lower fluxes across a range of frequencies. In the region of 10~MHz to 100~GHz, the wind fluxes range from $\sim10^{-3}$ to $5$~$\mu$Jy. Therefore, the peak flux from the wind emission is four orders of magnitude smaller than the peak flux of the planet. As a result, it should be straightforward to distinguish between the planetary and wind emission.

\begin{figure}
\centering
\includegraphics[width = \columnwidth]{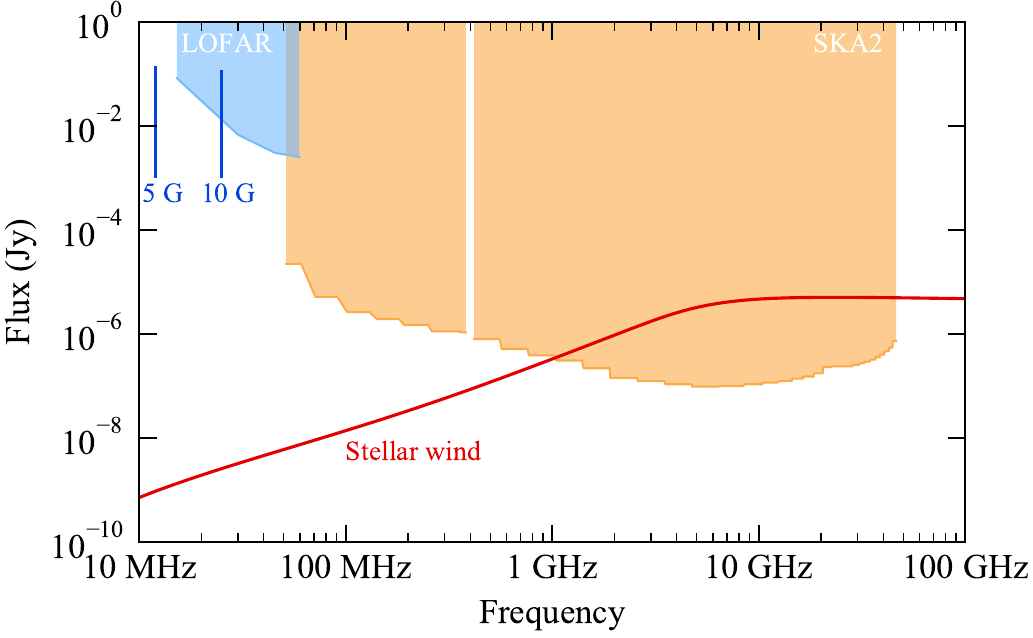}
\caption{Comparison of the predicted radio flux densities and frequencies of HD189733b, for assumed planetary field strengths of 5 and 10~G, and the stellar wind of its host star. The peak planetary flux density shown is the max over the three epochs for a given field strength, with a lower limit set of 1~mJy. The stellar wind spectrum is the same as shown in Figure~\ref{fig:wind radio spectrum}. The sensitivities of LOFAR and SKA2 for a 1~hour integration time are also shown, adapted from \citet{griessmeier11} and \citet{pope18} respectively.}
\label{fig:radio sensitivites}
\end{figure}

Figure~\ref{fig:radio sensitivites} shows a comparison between the predicted fluxes from the planet and wind of the host star, along with the sensitivities of LOFAR and SKA2. We see that for a planetary field strength of 5 G, the corresponding cyclotron frequency of 12~MHz is below the lower frequency limit of LOFAR quoted by \citet{griessmeier11}. However, we determined in Section~\ref{sec:propagation of the planetary radio emission} that propagation of emission below 21~MHz is unlikely. For the stellar wind, we see that SKA2 is likely to have sufficient sensitivity to determine at which frequency it becomes optically thin. If so, this would allow for the base density of the stellar wind $n_0$ to be constrained, and consequently the mass-loss rate of the star (see Appendix \ref{sec:appendix - wind radio emission lower density}).

We note that there are likely additional sources of radio emission in planetary system, such as thermal emission from the stellar chromosphere \citep{villadsen14, fichtinger17}. However, we do not investigate this in this work.

% Planet and wind fluxes and frequencies with telescope sensitivities
% #########################################################
%  Conclusions
% #########################################################

\section{Summary \& Conclusions}
\label{sec:conclusions}

In this work we have characterised the radio environment of the hot Jupiter HD189733b. For its host star HD189733, we performed 3D MHD simulations of its wind, implementing surface magnetic field maps at the epochs Jun/Jul~2013, Sep~2014, and Jul~2015, which were reconstructed from observations. We used our models to calculate the wind variability of HD189733, and determined that the mass-loss rate, angular momentum-loss rate, and open magnetic flux vary by 9\%, 40\%, and 19\% over this period respectively. We also found from our wind models that the planet experiences a non-uniform wind as it progresses through its orbit, with the wind velocity and particle number density varying by 29-37\% and 25-32\% respectively over the three modelled epochs. Temporal variations were observed in the extended atmosphere of HD189733b \citep{lecavelier12}, which could arise from interactions with a variable stellar wind \citep{bourrier13}. Our derivation of the stellar wind properties at the planetary orbit for different epochs will thus be useful to the interpretation of our MOVES observations of the upper planetary atmosphere.

Using the numerical code developed by \citet{ofionnagain19}, we calculated the free-free radio spectrum of the wind of HD189733. We found that it emits at low fluxes of $10^{-3}$ to $5$~$\mu$Jy in the frequency range of 10~MHz to 100~GHz. While it is unlikely that this could be detected with current radio telescopes, future endeavours such as SKA are likely to have the sensitivity required to characterise the radio emission from the winds of stars such as HD189733. If detectable, the mass-loss rate of star could be constrained.

Our wind models provided us with the local wind properties at the planetary orbit, which in turn we used to calculate the planetary radio emission. For assumed planetary field strengths of 1, 5, and 10~G respectively, we found that the planet emits at frequencies of 2, 12, and 25~MHz, with little variation in these values over the modelled period and the orbit of the planet. The emitted flux density however varies from 1 to $10^2$~mJy in our calculations. Therefore, a frequency corresponding to a field strength $\gtrsim 10$~G places HD189733b within the detection limit of LOFAR.

We investigated if our predicted planetary radio emission could propagate, and found that HD189733b orbits in and out of regions where the planetary emission will be absorbed by the stellar wind. For the calculated cyclotron frequencies of 12 and 25~MHz, we found that propagation can only occur for 41\% and 67\% of the orbit respectively. The fraction of the orbit where propagation is possible corresponds to when the planet is approaching and leaving primary transit of the host star. This could be useful information in planning radio observing campaigns of exoplanetary systems. However, as the radio flux densities of the planet are several orders of magnitude higher than those of the stellar wind, some of the planetary emission may still escape even after being attenuated in the stellar wind. We also determined that the plasma frequency of the stellar wind at the planetary orbit is too high for propagation to occur below 21~MHz at the three modelled epochs. In addition to this, the planetary atmosphere itself may prevent the generation of radio emission, as has been recently suggested by \citet{weber18}.

To conclude, our work has shown that the most favourable candidates for detection of planetary radio emission are hot Jupiters with large magnetic field strengths, orbiting inactive stars with low density winds.

% #########################################################
%   Acknowledgements
% #########################################################

\section*{Acknowledgements}

The authors thank the anonymous referee for their comments and suggestions. RDK acknowledges funding received from the Irish Research Council through the Government of Ireland Postgraduate Scholarship Programme. RDK and AAV also acknowledge funding received from the Irish Research Council Laureate Awards 2017/2018. 
VB acknowledges support by the Swiss National Science Foundation (SNSF) in the frame of the National Centre for Competence in Research PlanetS, and has received funding from the European Research Council (ERC) under the European Union's Horizon 2020 research and innovation programme (project Four Aces; grant agreement No 724427). This work was carried out using the BATSRUS tools developed at The University of Michigan Center for Space Environment Modeling (CSEM) and made available through the NASA Community Coordinated Modeling Center (CCMC). The authors also wish to acknowledge the SFI/HEA Irish Centre for High-End Computing (ICHEC) for the provision of computational facilities and support.

% #########################################################
%  Bibliography
% #########################################################

\bibliographystyle{mnras}
\bibliography{bibliography}

% #########################################################
%   Appendix - Effects of a lower wind density on the wind radio emission
% #########################################################

\appendix

\section{Radio emission and absorption for a lower density stellar wind}
\label{sec:appendix - wind radio emission lower density}

The free-free radio emission spectrum of the stellar wind depends on the density. The density profile is primarily determined by the coronal base density in our wind simulations $n_0$, which is a free variable. To see the effect of a lower density wind on the free-free radio spectrum of HD189733, we divide our density values from the wind simulations by a factor of 10. 

\begin{figure}
\centering
\includegraphics[width = \columnwidth]{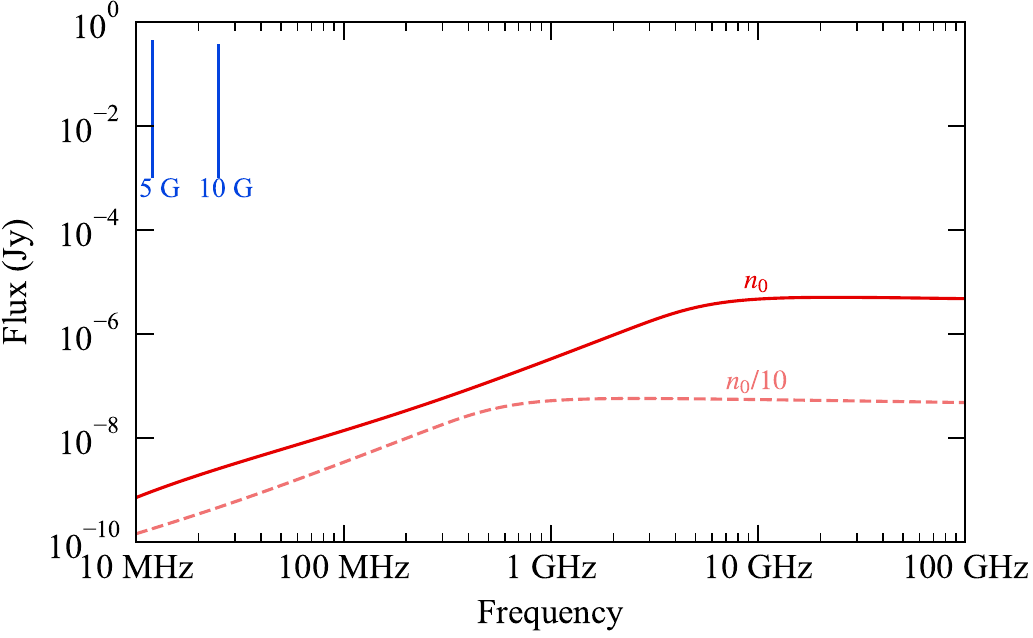}
\caption{Comparison of the free-free radio spectrum of the stellar wind of HD189733 from Figure~\ref{fig:wind radio spectrum} ($n_0$), with a stellar wind that has a density which is a factor of 10 lower ($n_0/10$). The planetary flux densities for assumed field strengths of 5 and 10~G are also shown in the upper left corner. The peak values of the planetary fluxes have been scaled for the lower density stellar wind from those shown in Figure~\ref{fig:radio sensitivites}.}
\label{fig:wind radio spectrum lower density}
\end{figure}

Figure~\ref{fig:wind radio spectrum lower density} shows the comparison of the spectrum shown in Figure~\ref{fig:wind radio spectrum} and the spectrum calculated for the lower base density wind. We find that the lower density wind becomes optically thin at 1~GHz, as opposed to 10~GHz for the higher density wind. We also see that for the lower density wind, the radio flux densities are 1-2 orders of magnitude smaller that those for the $n_0 = 10^{10}$~cm$^{-3}$ wind. So, detection of free-free emission from the stellar wind of HD189733 is more favourable if it has a denser wind. However, the fluxes calculated for the higher density wind are still quite small. While it is unlikely that this could be detected with current radio telescopes, it is expected that future developments such as SKA will allow for radio emission from the winds of nearby low mass stars to be detected \citep[see][]{ofionnagain19}. By determining the frequency at which the wind becomes optically thin, the mass-loss rate of the star could be constrained.

We also show the planetary flux densities for 5 and 10~G in Figure~\ref{fig:wind radio spectrum lower density}, which have been scaled for the lower density stellar wind. Assuming the planet remains orbiting in the ram-pressure dominated regime for the lower density wind, the planetary radio flux scales with ${n_0}^{-1/2}$ \citep{vidotto17a}. The frequency of the planetary emission has a very small dependence on the stellar wind density, as the magnetopause size \Rm\ scales with ${n_0}^{-1/6}$ in this regime (see Equations~\ref{eq:magnetopause size} to \ref{eqn:cyclotron frequency planet}).

\begin{figure}
\centering
\includegraphics[width = \columnwidth]{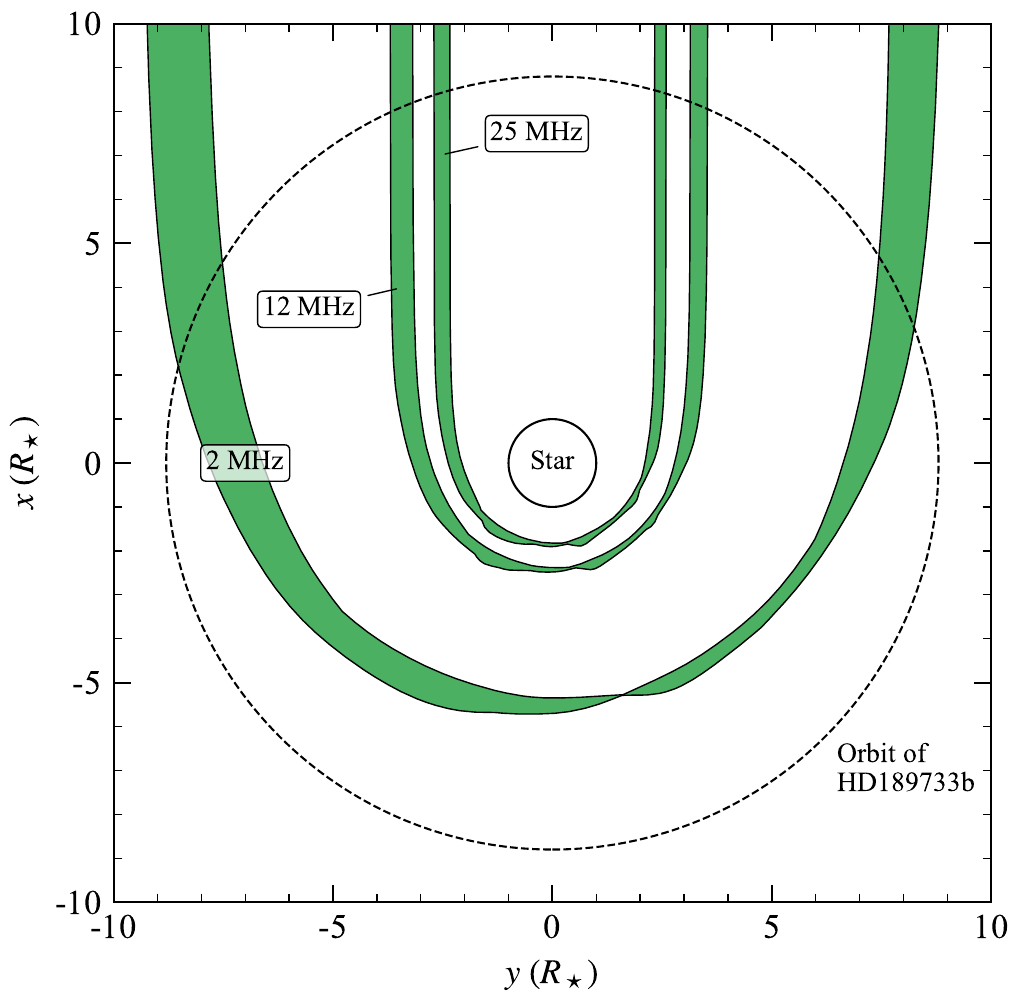}
\caption{Radio photospheres of the stellar wind of HD189733, for a stellar wind that is a factor of 10 lower in density. The sizes of the radio photospheres of the lower density wind at a given frequency are smaller compared to those shown in Figure~\ref{fig:radio photospheres}.}
\label{fig:radio photospheres lower density}
\end{figure}

The sizes of the radio photospheres of the wind will also depend on the wind density. Figure~\ref{fig:radio photospheres lower density} shows the boundaries of these regions at the three calculated cyclotron frequencies in the orbital plane of the planet for the lower density wind. We see that for a lower density, the regions the planet orbits through where wind is optically thick to the cyclotron frequencies are smaller compared to those shown in Figure~\ref{fig:radio photospheres} at a given frequency. Therefore, contrary to the wind radio emission, detection of planetary radio emission is more favourable for planets orbiting stars with low density winds.

% Free-free wind radio spectrum for lowered density

% Photospheres in the orbital plane for lowered density

% #########################################################
%  End of document
% #########################################################

% Typesetting comment
\bsp
\label{lastpage}
\end{document}